\documentclass[prl,manuscript,aps]{revtex4}
\usepackage{amssymb,amsmath}
\usepackage{graphicx,float}

\begin{document}

\title{ \Large \bf \flushleft 
Fractal noise maps reveal human brain activity: 
a key for unbiased fMRI analysis}
\author{\large  \flushleft 
Stefan Thurner$^{1}$\footnote[15]{
 {\bf Correspondence to:}\\
 Prof. Stefan Thurner PhD, PhD;  HNO, AKH-Wien, University of Vienna\\
 W\"ahringer G\"urtel 18-20; A-1090 Vienna, Austria\\
 Tel.: ++43 1 40400 2099;  Fax: ++43 1 40400 3332\\
 e-mail: thurner@univie.ac.at
}, 
Christian Windischberger$^{2}$, 
Ewald Moser$^{2}$,   
Markus Barth$^{2}$ }

\affiliation{ \flushleft
$^1$Institut f\"ur Mathematik NuHAG and Universit\"atsklinik f\"ur HNO,    
$^2$Universit\"atsklinik  f\"ur Radiodiagnostik,
University of Vienna; W\"ahringer G\"urtel 18-20;    
A-1090 Vienna;  Austria  \\ \hspace{1cm}
\\} 

\date{August 2002}
\begin{abstract}
\noindent
Brain metabolism is controlled by complex regulation mechanisms.
As part of their nature many complex systems show scaling behavior 
in their timeseries data. Corresponding scaling exponents can sometimes 
be used to characterize these systems. Here we present maps of 
scaling exponents derived from BOLD functional Magnetic Resonance 
Imaging (fMRI) data, and show that they reveal activation patterns 
in the human brain with very high precision. In contrast to standard 
model-based analysis we use no prior knowledge on the experimental 
stimulation paradigm for extracting activation patterns from fMRI 
timeseries. We demonstrate that mental activity is one-to-one 
related to large fractal exponents, or equivalently, to 
temporally highly correlated processes.

\vspace{1cm}
\noindent
PACS:
87.10.+e, 
87.57.-s, 
87.61.-c, 
05.45.Df, 
05.45.Tp  
\end{abstract}

\maketitle

Functional magnetic resonance imaging (fMRI) 
based on Blood Oxygenation Level Dependent  (BOLD) 
signal-changes allows assessment of brain activity
via local haemodynamic variations over time 
\cite{ogawa93}.  
It provides the highest combined spatial
and temporal resolutions presently available for  
non-invasive functional brain mapping. 
In fMRI measurements slices of the human brain are
sampled in a repetitive manner with mm$^3$ sized 
voxels as the smallest information-carrying spatial element.
These represent aggregated and collective information on
the physiological status of a large number 
of neurons with a temporal resolution of tens of ms. 
In a typical fMRI experiment external (e.g. visual) stimuli  are 
presented in intervals of several seconds ($\sim$ 0.1 Hz), 
causing a change in voxel-signal intensity, delayed and blurred 
by the haemodynamic response  
\cite{Aguirre98,Goutte01}.
Functional MRI voxel-timeseries 
exhibit considerable noise components which adversely
effect conventional fMRI analysis and  
therefore are typically considered undesired and interfering. 
On the other hand, by studying fMRI noise evidence has been  found
that voxel-timeseries potentially exhibit 1/f noise characteristics 
\cite{zarahn97,steve}. 
1/f noise -- e.g. characterized by a power-law decay of the 
Power Spectrum Density ($PSD$) --  has been discovered in 
a wide variety of systems which are governed by 
complex physiological regulation mechanisms such as the 
cardiovascular system 
\cite{kobayashi82,teich96}. 

Brain metabolism is controlled by a variety of complex 
mechanisms that guarantee proper brain function including   
sufficient blood supply to meet neural demands. 
Increased metabolic demand is triggered by mental processes
and initiates a cascade of biochemical reactions,   
followed by changes in haemodynamic parameters such as 
blood flow and oxygenation 
\cite{nature01}. 
In this work we argue that the use of high 
temporal-resolution BOLD fMRI 
in combination with a scaling analysis
enables us to distinguish different physiological 
states (active and non-active) of the brain. 
We show that this is a robust method and can be seen already by 
using relatively simple fractal noise-measures 
like Hurst- and $PSD$-exponents. 
We focus on the statistical scaling 
structure of voxel-timeseries obtained in fMRI experiments 
and visualize the amount of information  
present in the ``noise''.

Noise structure can be characterized by so-called fractal measures. 
Usually one uses more than one method to avoid well known 
systematic deficiencies inherent to either method 
\cite{thurner97}. Here we use two of the simplest methods available.  

{\it Hurst Exponent:} 
The concept of Brownian motion (random walk) 
can be generalized to fractional Brownian motion \cite{mandelbrot68}
by introducing a scaling exponent $H$, often called Hurst exponent.
Maybe the most transparent way to define it is by the two-point 
correlation function of a (stochastic) data timeseries $x(t)$, 
\begin{equation}
 C(\tau) =  \langle [ x(t+\tau) - x(t)]^2 \rangle \propto \tau^{2H}  \quad, 
\label{hurst}
\end{equation}
with $H$ being a real number, $0<H<1$. 
$H$ serves to characterize the process in terms of temporal correlations: 
while for classical Brownian motion ($H=0.5$) there are no correlations, for 
$H>0.5$ the process is called positively correlated or persistent, 
i.e., if the process was moving upward (downward) at 
time $t$ it will tend to 
continue to move upward (downward) at future times $t'>t$ as well. 
This means that increasing (decreasing) trends in the past 
imply -- on average -- 
increasing (decreasing) trends in the future 
\cite{feder88}. 
If  $H<0.5$  the corresponding process is called anti-correlated 
or anti-persistent, meaning that increasing (decreasing) trends in 
the past imply -- on average -- decreasing (increasing) trends. 

{\it Power-Spectrum exponent:}
Alternatively, one can quantify internal noise structure 
in stochastic processes by the power-spectrum exponent $S$, 
defined by the power characterizing the decay of the (discrete)
$PSD$, i.e., the squared Fourier spectrum 
of the process $x(t)$, 
\begin{equation}
 PSD(\omega)= \left| \sum_{k=1}^{N} x(k) e^{i 2 \pi (\omega-1)(k-1)/N }  
\right| ^2  
\propto \omega ^{-S} \quad,
\label{psd}
\end{equation}
with $N$ being data size.
For perfect 1-dimensional fractal processes it can be shown
that $H$ and $S$ are related by $S=2H+1$.
Hence, exponents $H$ ($S$) different from a value of  
0.5 (2) indicate a self affine, internal structure 
of the timeseries. 

Functional MRI was performed on a 3 Tesla whole-body MR scanner
(MedSpec S300, Bruker, Ettlingen, Germany)
using  single-shot gradient echo--planar imaging (EPI),  
with a $T_R$ of $200$ ms and an effective $T_E$ of $28$ ms. 
Four 4~mm thick slices parallel to the calcarine sulcus were 
measured using a field-of-view of $29\times29 \, {\rm cm}^2$ 
and a matrix size of $64\times64$ pixels.
This setup was repeated 1500 times resulting in a total 
measurement time of 5 minutes.
A visual stimulation paradigm was presented at randomly 
distributed time points during scanning, separated by an 
interval of at least 10 seconds, 
guaranteeing that all stimulation frequencies are below $0.1$ Hz. 
Stimulation consisted in presenting a rotating 
checker-board disk to the subjects for three seconds 
via video projector and mirror.

Voxel time-course information $\bar I(\vec x,t)$ was 
obtained from images taken at time instances $t$ ($\vec x$ 
is voxel-position in 3D space).  
For further analysis we subtracted the 
temporal mean from $\bar I(\vec x,t)$ to obtain the process 
$I(\vec x,t)= \bar I(\vec x,t)-\langle \bar I(\vec x,t)\rangle_t$. 
This process serves as the starting point for all further
analysis, no more pre-processing steps,  
neither filtering, normalization, nor motion 
correction were applied to the data.
For fractal analysis we use the cumulative voxel-intensity process
$x(\vec x,t)= \sum_{i=1}^{t} I(\vec x,i)$.

Standard fMRI analysis of the data sets was performed 
using linear regression, 
based on the known stimulation paradigm convolved with a model 
haemodynamic response function, resulting in a model 
function $HR(t)$ \cite{Friston95}.
The results of this analysis are r-maps, i.e. correlation coefficients 
of the model function $HR(t)$ with actual fMRI data for all individual 
pixels at positions $\vec x$. 
Results are thresholded to reveal the pixels activated during visual 
stimulation. In the following a t-threshold of 5 was chosen 
which corresponds to a Bonferroni-corrected $p$-value of 0.001, 
representing highly significant activation.

Fig. \ref{Fig_intensity} displays the averaged time-course
of all activated pixels of a single, 4-slice experiment as 
found by standard regression analysis \cite{SPM}.  
The corresponding functional r-maps in Fig. \ref{Fig_res1}a
clearly show the
expected functional activation in the primary visual areas.  
The respective time-course is in excellent agreement 
with the stimulus applied,  as indicated by the red bars 
in Fig. \ref{Fig_intensity}. 
In Fig.~\ref{Fig_res1}b and c, maps of 
$H$ and $S$ values are shown  
for all voxels within the brain slices, respectively. 
The color table is the same for all three parameters,
however, adjusted to match the respective range of values. 
Regions with a large Hurst-value (red)
coincide with areas of visual activation,
as detected by the regression analysis
with high statistical significance ($p<0.001$). 
The same regions can be found in the $S$-maps, however, 
to somewhat less spatial precision.
Signals in frontal regions are caused by movement artifacts.
We checked this by confirming that they reduce when applying 
standard realignment software.  

To demonstrate that the observed changes in the scaling 
parameters $H$ and $S$ are in fact properties of noise,
we performed the following checks: 
{\it (i) Residual analysis}: We pixel-wise fitted our data 
to results of the model function $HR(\vec x,t)$ obtained from a 
linear standard fMRI model \cite{Friston95,SPM},  
\begin{equation}
\bar I(\vec x,t)= HR(\vec x,t) + \xi(\vec x,t) \quad , 
\end{equation}
where $\bar I(\vec x,t)$ is the fMRI signal 
and  $\xi(\vec x,t)$ the residual noise at pixel position $\vec x$. 
We performed the same scaling exponent analysis as before 
starting from the residual dataset $\xi(\vec x,t)$. 
As expected the correlation coefficients become vanishingly small and 
can not serve as a reliable measure anymore,  Fig. \ref{Fig_res2}a.  
With $\xi(\vec x,t)$ as the input, $H$ and $S$ are shown in 
Fig. \ref{Fig_res2}b and c. The same areas of activity are recovered.
This is a clear sign that $\xi(\vec x,t)$ which is often (wrongly) 
considered ''white noise'' still carries relevant information. 
One could argue here that if this is the case the model is not yet 
good enough and has to be improved before discussing statistical 
significance of activation patterns. 
However, actually used models are based on linear approaches 
and can not account for the scaling effects reported here. 
{\it (ii) Low pass filtering}: By applying successive temporal smoothing of 
the pixel timeseries we find that the scaling effects 
eventually disappear, and obviously the structure observed here 
is a characteristics of the noise and not of the low frequency stimulation 
(signal). 
{\it (iii) Background noise is white}: 
We find that the background noise (measurement or machine errors)
in the signal as observed outside the head fluctuates 
around $H=0.5$ (mean and standard deviation: $H = 0.492 \pm 0.046$), 
as expected. 

In Fig. \ref{methode} the timeseries $I(\vec x,t)$ for two selected voxels 
are shown,
identified as either non-activated (a) or activated (b) 
in the standard fMRI analysis.
Periods of visual stimulation are again marked with red bars.
Clearly, the voxel in the visual region follows 
the stimulus (much like the entire cluster; 
compare Fig. \ref{Fig_intensity}).
In Fig. \ref{methode}c and d we show  the 
$PSD$ of $x(\vec x,t)$.
Voxel timeseries show typical 1/f noise characteristics and   
the power spectral density exponent $S$ can be estimated safely
by fits to the $PSD$-slope in a log-log plot. 
Fits are given by red lines. 
In order to focus exclusively on the noise structure and to 
exclude any influence of the BOLD response itself,
the frequency range corresponding to the stimulus presentation
(between 0.04 - 0.1 Hz) was excluded from the fit-range.
Frequency fluctuations larger than 1 Hz were also
not included as those frequencies are above the fractal onset frequency
for the given data size \cite{thurner97}. 
In Fig. \ref{methode}e and f we plot $C(\tau)$ from Eq. \ref{hurst}, 
and extract the Hurst exponent $H$ as half of the slope of the linear fit. 
Again, the same effective fit-range (red lines) was used.

A priori, noise components above the characteristic stimulus 
frequencies could be thought to appear as 
measurement noise and undesirable artifacts. 
In this work we have examined these noise components from 
a scaling point of view. 
We found that in non-active cortical regions, voxel-activity 
is well described by classical Brownian motion (random walk model,  
$H \sim 0.5$ and $S \sim2$), while noise components 
from voxels of active brain regions 
-- besides  obviously following the stimulus -- 
change their character towards highly correlated stochastic processes, 
i.e., correlated fractional Brownian noise. 
We have also observed that the latter regions are detected by $H$ and $S$ 
at very high sensitivity, i.e.,   
regions of large $H$ and closely correlate with
large r-values from model-based regression analysis. 
In areas outside the subject's head, $H$ is fluctuating 
around $0.5$ ensuring that no internal noise structure is 
arising or due to measurement or machine noise. 
These findings clearly demonstrate persistent fractal dominance of 
noise in voxel-timeseries in regions of neural 
activation, while in other areas
uncorrelated noise prevails,
consistent with recent work 
\cite{weisskoff,bullmore01,krueger01}.
$S$ and $H$ seem to be related to some extent, 
indicating the voxel-processes to be close to perfect fractals. 

We find that even residual timeseries, as e.g. obtained by
SPM software \cite{SPM}, show practically the same scaling information content. 
This finding has far reaching consequences for statistical 
analyses of fMRI data. Standard statistics is based on models which 
leave white noise as a residuum. Based on white noise,  
null hypotheses can be tested and sensible $p$-values be quoted.   
For correlated or scaling processes, however, it is known that standard 
statistics is not valid, since residua are by definition 
correlated (e.g. via scaling relations).
As a consequence other methods based on more complicated statistics 
have to be used to arrive at ''$p$-values'' which capture actual statistical
significance. Maybe Tsallis entropy \cite{tsallis} and its associated 
non-extensive statistics 
offer a natural starting point for consistently solving this problem. 

In conclusion we find that the statistical nature of 
voxel-timeseries depends on the underlying level of neural activity.
In activated regions the noise processes are 
strongly temporally correlated. 
The practical potential of this study is that scaling analysis is entirely 
model-independent: we do not make any assumptions about 
the nature of the voxel-processes, but rather try to 
clarify it.
Moreover, no information on the stimulus paradigm is required. 
It is intriguing that the information carried by 
fMRI noise alone is sufficient to at least compete with the 
gold-standard of fMRI analysis for detecting statistically 
significant regions of neural activation.  
Scaling measures -- as a complementary source of information --
may therefore improve  specificity in fMRI data analysis.  
We believe that the use of fractal parameters in fMRI also represents
a significant potential for understanding the
fundamental processes involved in neuro-vascular coupling, 
by identifying ``noise components'' as valuable measurements representing 
detailed high(er) frequency information, rather than treating 
``noise'' as a nuisance.  Our results can be seen in the context 
of recent trends which focus on the practical role of noise \cite{SR}, 
and that one should not restrict oneself to the obvious, 
but also hunt for the information hidden in the ``noise''. 


\newpage 
\bibliographystyle{unsrt}

\begin{thebibliography}{99}
\bibitem{ogawa93}
S. Ogawa, R.S. Menon, D.W. Tank, S.G. Kim, H. Merkle, J.M. Ellerman, K. Ugurbil,
{J. Biophys.} {\bf 64}, 803­812 (1993).  

\bibitem{Aguirre98} 
G.K. Aguirre, E.  Zarahn, M.  D'Esposito,  
{NeuroImage} {\bf 8}, 360-369 (1998). 

\bibitem{Goutte01}
C. Goutte, L.K. Hansen, M.G. Liptrop, E. Rostrup,  
{Human Brain Mapping} {\bf 13}, 165-183 (2001).

\bibitem{zarahn97}
E. Zarahn, G.K. Aguirre, M. D'Esposito,  
{NeuroImage} {\bf 5}, 179-197 (1997). 

\bibitem{steve}
S.B. Lowen, private communication. 

\bibitem{kobayashi82} 
M. Kobayashi, T.  Musha,  
{IEEE Trans. Biomed. Eng. } 
{\bf 29}, 456-457 (1982).

\bibitem{teich96} 
R.G. Turcott, M.C. Teich,   
{Ann. Biomed. Eng.} {\bf 24}, 269-293 (1996).

\bibitem{nature01}
N.K. Logothetis, J.  Pauls, M.  Augath,  T. Trinath,  A. Oeltermann,  
{Nature} {\bf 412}, 150-157  (2001).  

\bibitem{thurner97}
S. Thurner, S.B.  Lowen, M.C. Feurstein, C.  Heneghan,  
H.G. Feichtinger,   M.C.  Teich,
{Fractals} {\bf 5}, 565-595 (1997).

\bibitem{mandelbrot68}
B.B. Mandelbrot, J.W.  van Ness,  
{SIAM Rev.} {\bf 10}, 422-437 (1968). 

\bibitem{feder88} 
J. Feder,  
{\it Fractals}, 
Plenum Press, New York (1988). 

\bibitem{Friston95} 
K.J. Friston, C.D. Frith, R.S. Frackowiak, R. Turner, 
{NeuroImage} {\bf 2}, 166-172 (1995).

\bibitem{SPM}
Statistical Parametric Mapping, see homepage
http://www.fil.ion.ucl.ac.uk/spm/ 

\bibitem{weisskoff}
R.M. Weisskoff, J. Baker, J. Belliveau, T.L. Davis, K.K. Kwong, 
M.S. Cohen, and B.R. Rosen, 
{Proceedings Int. Soc. for Magn. Reson. in Med.} {\bf 1}, 7 (1993). 

\bibitem{bullmore01}
E. Bullmore, {et al.}, 
{Human Brain Mapping} {\bf 12}, 61-78 (2001). 

\bibitem{krueger01}
G. Kr\"uger,  G.H.  Glover,  
{Magn. Reson. Med.} {\bf 46}, 631-637 (2001).

\bibitem{tsallis}
C. Tsallis, J. Stat. Phys. {\bf 52}, 479 (1988). 

\bibitem{SR} 
J.J. Collins,  
{Nature} {\bf 402}, 241 (1999).  
\end{thebibliography}

\newpage  
\begin{figure}[H]
\caption{Average time-course of 
all activated pixels (blue line).
Overlayed (red dashed line) is the functional paradigm
convolved with the haemodynamic response (model function $HR(t)$). 
The stimulation periods are indicated by red bars at the bottom.}
\label{Fig_intensity}
\end{figure} 

\begin{figure}[H]
\caption{Computed ``activation'' maps of  four brain slices using 
the standard linear regression analysis. Correlation coefficients 
(r-map) of the actual data and the functional paradigm
convolved with the haemodynamic response function are shown (a).  
Hurst exponent analysis for the pixel timeseries (b);
Power spectral exponent analysis for the pixel timeseries (c). 
}
\label{Fig_res1}
\end{figure}

\begin{figure}[H]
\caption{Same as previous figure, but starting from  
the residual dataset $\xi(\vec x,t)$.
Clearly the correlation map
(a) does not show activation patterns anymore, 
notice the change in scale compared to the previous figure. 
$H$ and $S$ maps, (b) and (c), show the same activated regions
and about the same numerical values as in the original data.} 
\label{Fig_res2}
\end{figure}

\begin{figure}[H]
\caption{Timeseries of a non-active (a) and an active (b) voxel.  
The power spectral exponent $S$ for these 
processes is given by the slope of the $PSD$ shown in 
(c) and (d). Clear 1/f behavior in the $PSD$ is 
observable. The Hurst exponent $H$ is defined as half the slope of the 
function $C$.}
\label{methode}
\end{figure} 

\newpage  
\begin{figure}[H]
\begin{center}

\vspace{4cm} 
\begin{tabular}{c}
\includegraphics[width=16cm]{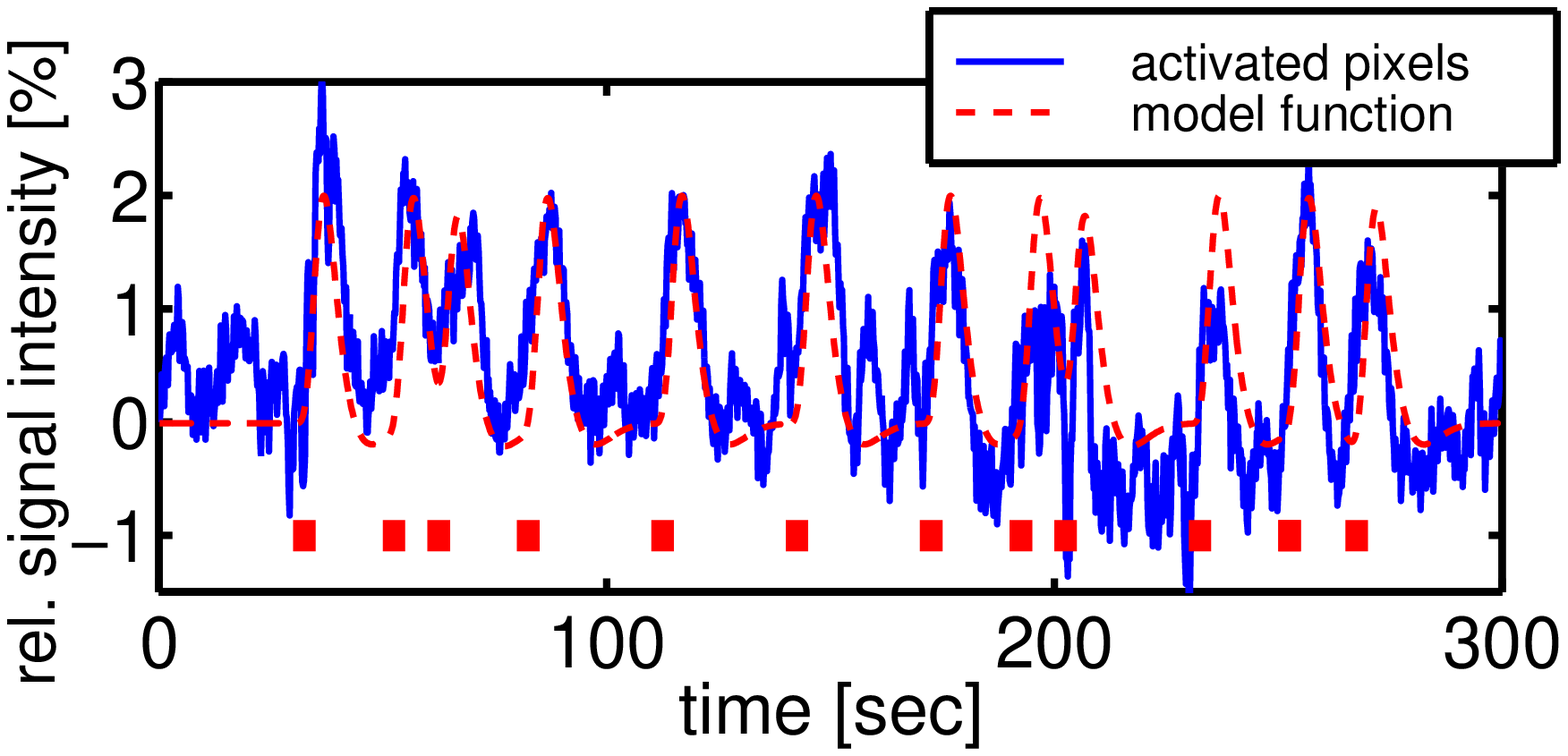}  \\

\end{tabular}
\end{center} 

\vspace{10cm} 
\begin{center} 
{\Large FIG. 1} 
\end{center} 
\end{figure} 

\newpage  
\begin{figure}[H]
\vspace{2.5cm} 
\hspace{-1.0cm} \includegraphics[width=19cm]{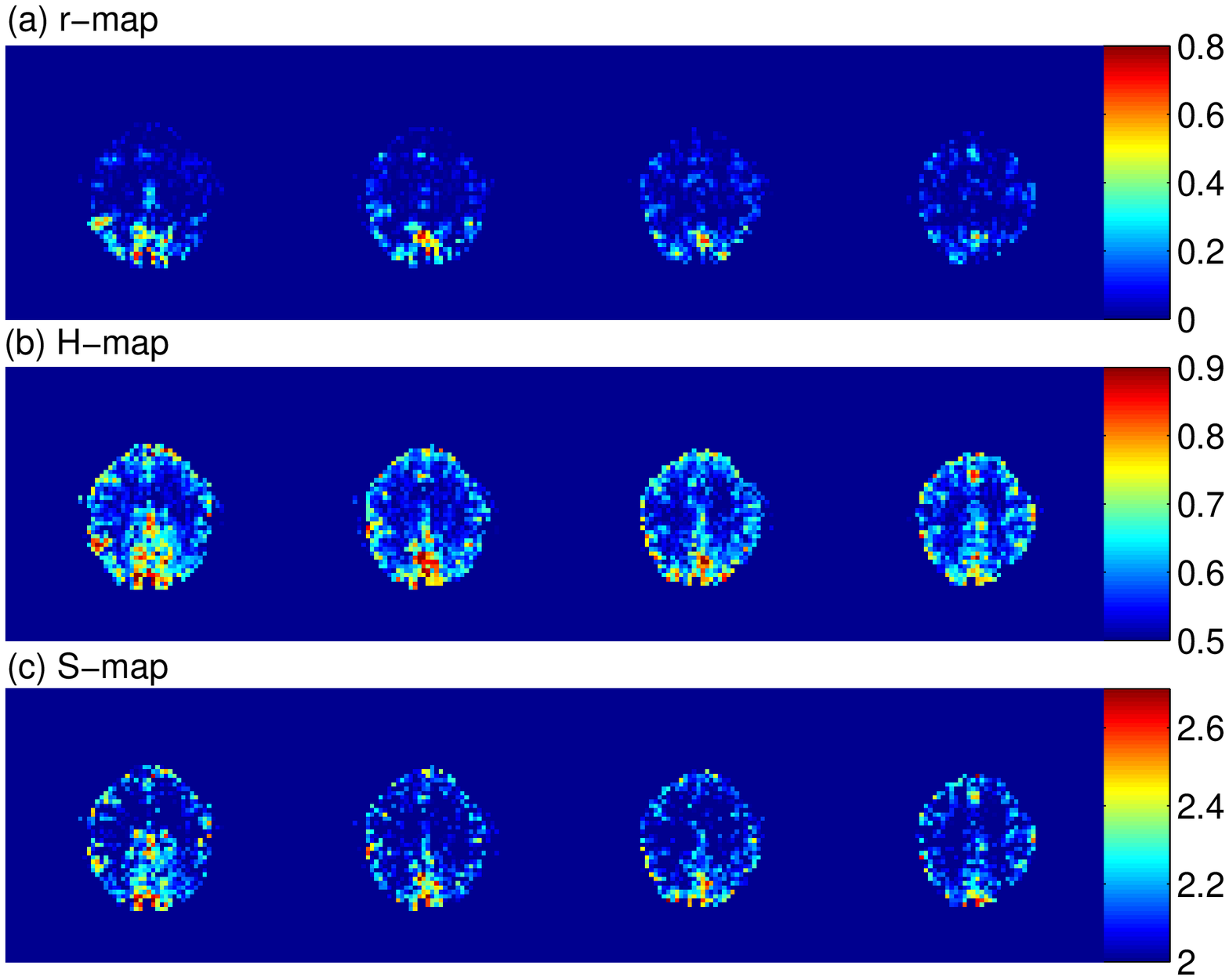} 

\vspace{4cm} 
\begin{center} 
{\Large FIG. 2} 
\end{center} 
\end{figure} 

\newpage  
\begin{figure}[H]
\vspace{2.5cm} 
\hspace{-1.0cm} \includegraphics[width=19cm]{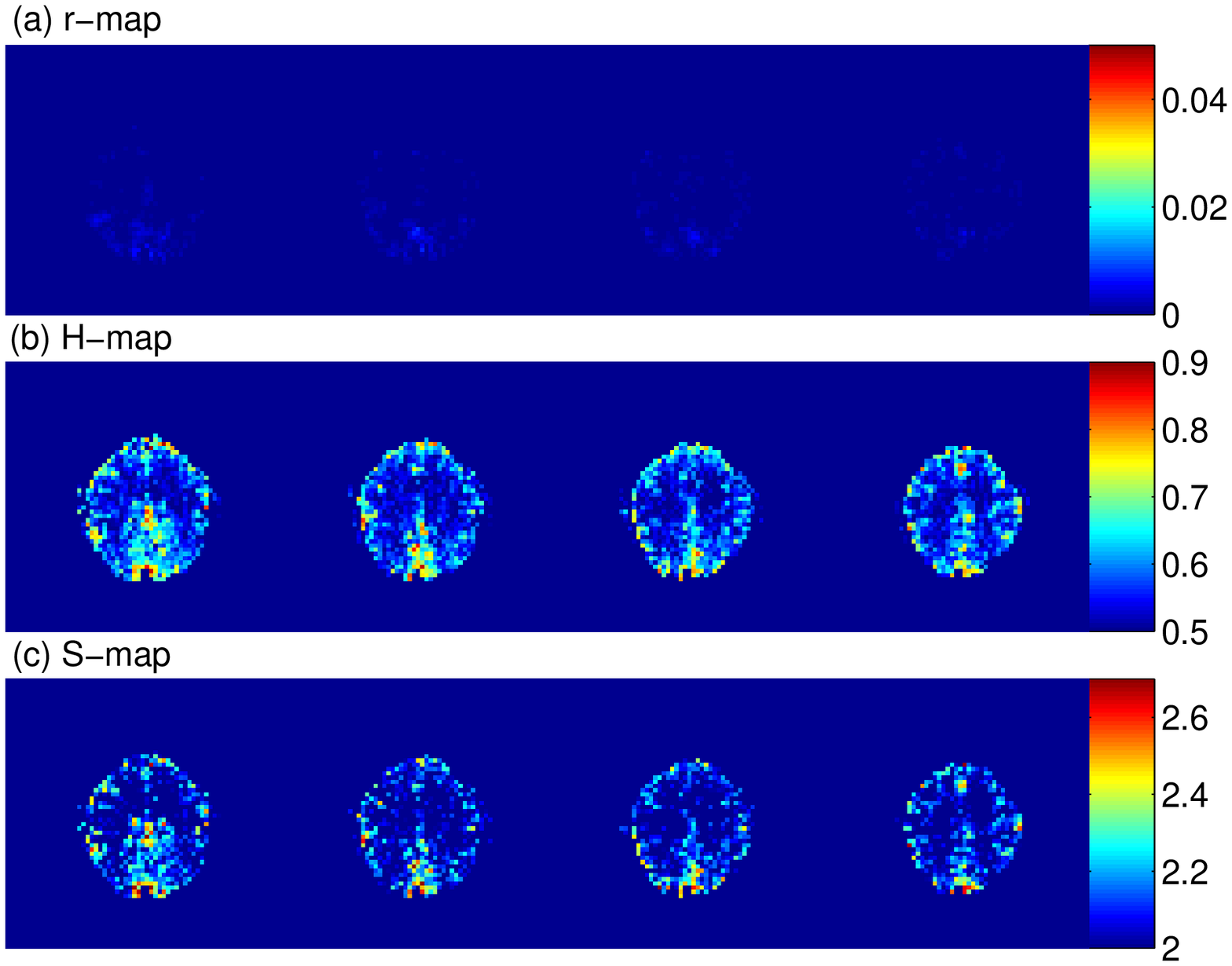} 

\vspace{4cm} 
\begin{center} 
{\Large FIG. 3} 
\end{center} 
\end{figure} 

\newpage  
\begin{figure}[H]
\begin{center}
\begin{tabular}{l}
\includegraphics[width=16.0cm]{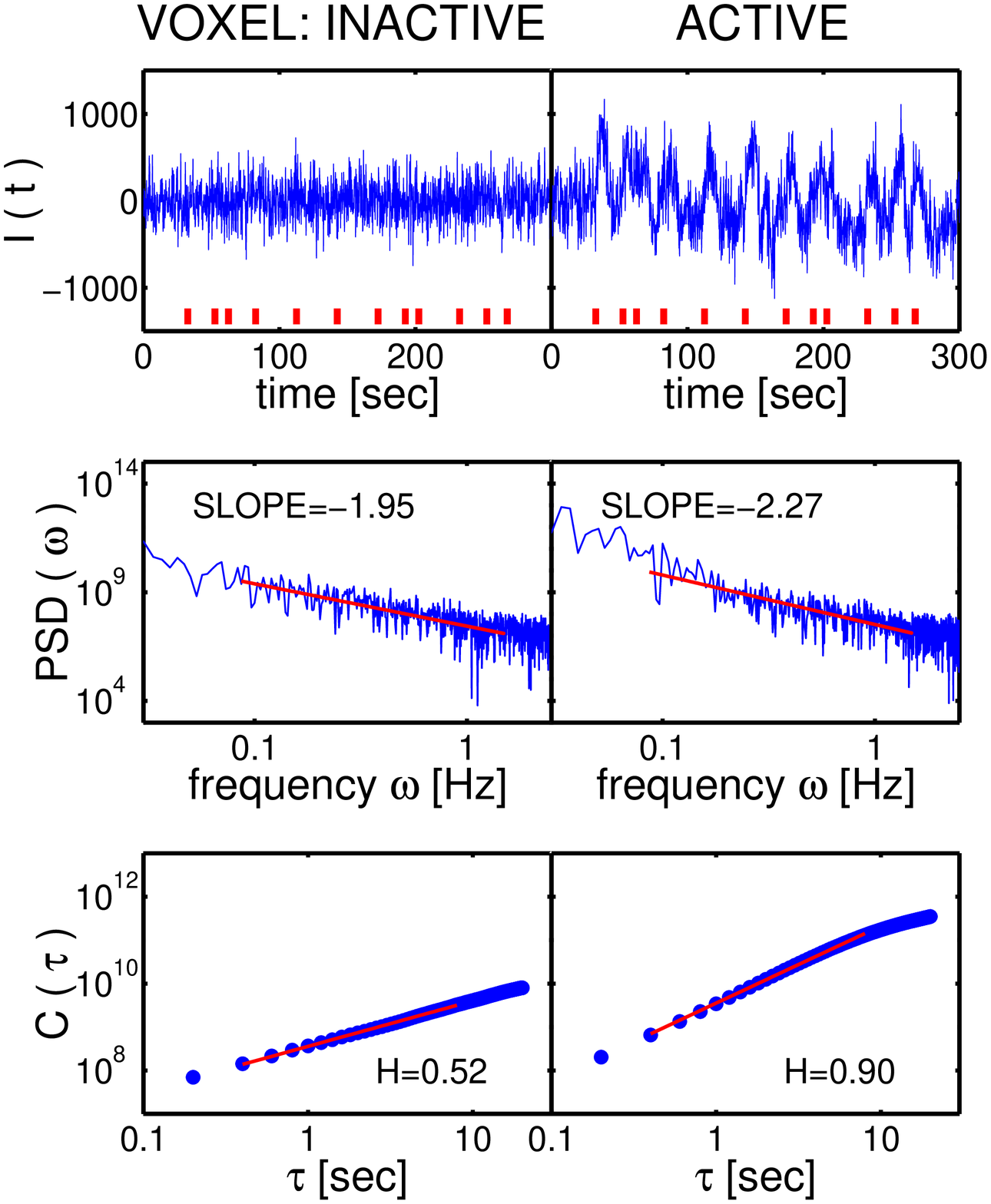}   \\
\end{tabular}

\vspace{-15.5cm}
\hspace{-3.5cm} {\LARGE (a) } \hspace{5.2cm} {\LARGE (b) } \\

\vspace{5.5cm}
\hspace{-3.5cm} {\LARGE (c) } \hspace{5.2cm} {\LARGE (d) } \\

\vspace{3.0cm}
\hspace{-3.5cm} {\LARGE (e) } \hspace{5.2cm} {\LARGE (f) } \\

\end{center} 

\vspace{6cm} 
\begin{center} 
{\Large FIG. 4} 
\end{center} 
\end{figure} 
%

\end{document}